\documentclass[aps,preprint]{revtex4}
\usepackage{amsfonts}%
\usepackage{amsmath}%
\usepackage{bm}
\usepackage{graphicx}
\usepackage{hyperref}

\begin{document}
\title{Relational time for systems of oscillators}
\author{G.J.Milburn and David Poulin}
\affiliation{The University of Queensland,  Department of Physics, School of Physical Science, Brisbane, Australia, }
\pacs{PACS number}

\begin{abstract}
Using an elementary example based on two simple harmonic oscillators, we show how a relational time may be defined that leads to an approximate Schr\"{o}dinger dynamics for subsystems, with corrections leading to an intrinsic decoherence in the energy eigenstates of the subsystem.  
\end{abstract}

\date{\today}
\maketitle

It is now well accepted that an energy superselection rule must be applied to the quantum description of the universe as a whole\cite{PW}. This implies that the universe does not evolve in coordinate time which is usually taken to mean that the quantum state is an energy eigenstate.  Quite how observers, internal to the universe, develop a notion of time and dynamics is then something of a puzzle\cite{Isham}. One solution is to try and make coherent the concept of an internal time based on special subsystems that fulfill the role of clocks. In a characteristically elegant presentation Peres\cite{Peres} has given a simple model for how this can be done. The Peres model, with three degrees of freedom, reduces to a Hamiltonian which is the difference of the Hamiltonians for two simple harmonic oscillators with the constraint that the total Hamiltonian vanishes. One oscillator can then be taken as a clock and the phase variable for this oscillator is taken as an internal time variable.  

In the example of this paper we consider a Hamiltonian which is the sum of two harmonic oscillator Hamiltonians. The constraint is imposed at the quantum level by required all physical states to be eigenstates of this Hamiltonian. We show how a relational view of time can be given using the fact that such eigenstates are simultaneous near eigenstates of the joint phase {\em difference} operator. This enables us to treat one oscillator as a clock. However quantum fluctuations in the  reduced state of the clock oscillator system lead to noise and complementary decoherence in the other oscillator. This is made explicit by introducing coordinate time using group averaging. A modified Schr\"{o}dinger equation is then obtained for the state of the system which explicitly includes decoherence in the energy basis. The approach taken here is similar to the fully relational approach described in \cite{Poulin05}.

While the standard approach takes the state of the universe to be an energy eigenstate, on information theoretical grounds, it is more reasonable to assign to the universe a mixed state which is diagonal in the energy basis as there is no way any observer inside the universe could know which particular eigenvalue is realised. Taking a Bayesian perspective, we need assume no more than that the quantum state of the universe is a mixture of energy eigenstates,
\begin{equation}
\rho_u=\sum_{\epsilon} p_{\epsilon} |\epsilon\rangle\langle \epsilon|.
\label{univ_state}
\end{equation}
At this level of description there is no way to assign the weights $p_\epsilon$, so we leave them arbitrary. 
 
 An alternative way to write the state in Eq.(\ref{univ_state}) is as a group average\cite{Marolf} over the one parameter unitary group generated by the Hamiltonian,
 \begin{equation}
 \rho_U=\lim_{T \rightarrow \infty}
\frac{1}{T}\int_0^T e^{-i\hat{H}t}|\Psi\rangle\langle \Psi|e^{i\hat{H}t} dt
\label{average}
 \end{equation}
 where $|\Psi\rangle=\sum_\epsilon c_{\epsilon}|\epsilon\rangle$
 is an arbitrary state. In this case $p_\epsilon=|c_\epsilon|^2$. 
In conventional quantum mechanics this is the group of time translations and the parameter is coordinate time. 

The universe is a rather large system and admits many decompositions into subsystems, some of which correspond to observers. One approach to recovering a notion of time is to partition the universe into at least two systems, one of which is to be taken as a {\em clock} and the rest, which we will simply refer to as the {\em system}.  It is then possible to decompose the state of the universe into a state in which a particular clock variable is seen to be highly correlated with particular states of the system. Under the right circumstances this correlation can be used to order the states of the system in such a way as to approximate the ordering imposed by the one parameter unitary group generated by the Hamiltonian of the system alone.   This is the so called relational view of time. As we shall see, the time evolution that results from a relational approach is not quite the same as unitary Hamiltonian evolution. Indeed, it is a special kind of non unitary process, defining a one parameter semigroup, and producing an unavoidable decoherence intrinsic to the quantum nature of the universe\cite{GPP2004,GM91}. 

As a specific example consider a completely closed system, a 'universe', composed of two simple harmonic oscillators. The total Hamiltonian is 
\begin{equation}
\hat{H}=\omega_1a_1^\dagger a_1+\omega_2a^\dagger_2 a_2
\label{tot_ham}
\end{equation}
(we work in units where $\hbar=1$). We have assumed that the oscillators do not interact. Anticipating the emergence of a relational time, we now take oscillator labelled $1$ as the {\em clock} and oscillator labelled $2$ as the {\em system}. 

We now impose a constraint on the allowed physical states of this system: the system is in a stationary state of the form Eq.(\ref{univ_state}) where $\epsilon$ is the total energy of both oscillators, that is to say, we restrict the physical Hilbert space to only those states that are eigenstates of the total Hamiltonian. 
In conventional quantum theory the joint state of two subsystems is described by the Hilbert space given by the tensor product of the Hilbert spaces for each subsystem. We can  use the eigenstates $|n\rangle_i$ of the number operators $a_i^\dagger a_i$ as a basis for each subsystem Hilbert space,
\begin{equation}
a_i^\dagger a_i|n\rangle_i=n|n \rangle_i,\ \ \ \ n=0,1,2,\ldots
\end{equation}
Thus a particular total energy eigenstate can be written as
\begin{equation}
|\epsilon\rangle=|n\rangle_1\otimes|m\rangle_2
\end{equation}
 with $\epsilon=\omega_1n+\omega_2m$. We now choose $\omega_1$ a unit of energy and write the total energy as 
 \begin{equation}
 \tilde{\epsilon}=n+\tilde{\omega}_2m
 \label{tot-energy}
 \end{equation}
 where $\tilde{\epsilon}=\epsilon/\omega_1,\ \tilde{\omega_2}=\omega_2/\omega_1$.
 
 For now let us simplify matters by assuming that $\omega_{1,2}$ and $\tilde{\epsilon}$ are commensurate. In particular we assume that 
 \begin{equation}
 \tilde{\omega}_2  =  N.
 \end{equation}
Together with Eq. (\ref{tot-energy}), this implies that $ \tilde{\epsilon}  =  M$ where $N,M$ are integers and $n,m$ are related by 
 \begin{equation}
 n = M-Nm.
 \end{equation}
Clearly the total energy eigenstate is $g(M)$-fold degenerate with $g(M) = \lfloor \frac MN \rfloor +1$.
We can then write a total energy eigenstate $|\epsilon\rangle\equiv|M:N\rangle$ as
\begin{equation}
|M:N\rangle=\sum_{m=0}^{g(M)-1} c_m |M-Nm\rangle_1\otimes|m\rangle_2
\label{energy_state}
\end{equation}
where, as we discuss below, the coefficients $c_m$ should be chosen roughly equal to approximately recover the conventional time evolution of the system. We note in passing that the coherence between the states $|M-Nm\rangle_1\otimes|m\rangle_2
$ superposed in Eq.(\ref{energy_state}) is immune to overall unitary transformations generated by the total Hamiltonian and is an example of a decoherence free state for this kind of collective noise, central to the fully relational construction of Ref.~\cite{Poulin05}. 

We regard the integer $M$ as a label for a particular degeneracy subspace, while the integer $N$ is ratio of the system (oscillator-$2$) frequency to the frequency of the clock (oscillator-$1$). In other words $N$ is the parameter that specifies the kind of two oscillator universe we are dealing with.  The state of this rather sparse universe may then be assigned as the mixture
\begin{equation}
\rho_N=\sum_Mp_M|M:N\rangle\langle M:N|
\label{mix-state}
\end{equation}
where $p_M$ is arbitrary at this stage. The assignment of a mixture rather than an energy eigenstate is consistent with the idea of a relational universe and follows from the idea that the particular energy realised by the universe cannot be known.  A direct application of  Bayesian reasoning would then suggests that the most appropriate state of the universe is the mixture, Eq.~({\ref{mix-state}). This of course is quite different to the approach taken in quantising ADM canonical gravity in which the Wheeler-de Witt equation implies that the universe is in a zero energy eigenstate\cite{Isham}.  However we expect that the more general statistical assumption can consistently be incorporated into that approach.  

As above, the one parameter group average Eq.~(\ref{average}) can be used to relate the coefficients $c_m$ and $p_M$ to an arbitrary state of the two oscillators $|\Psi\rangle = \sum_{nm} \alpha_{n,m} |n\rangle_1\otimes |m\rangle_2$. Indeed, we get
\begin{equation}
p_M = \left| \sum_{m=0}^{g(M)-1} \alpha_{M-Nm,m} \right|^2 \quad {\rm and}\quad 
c_m = \frac{\alpha_{M-Nm,m}}{\sqrt{p_M}}.
\end{equation} 

We now turn to physical quantities that we are permitted to use to describe this universe, that is those represented by operators that commute with the total Hamiltonian for the double oscillator system as only such operators will be independent of coordinate time. The only local operators (that is acting in the individual subsystem Hilbert spaces) that satisfy this constraint are functions of the local number operators, $a_i^\dagger a_i$. However when we turn to joint operators the situation is more interesting.  As we now show the phase difference operator is an important example.

Define the canonical unitary displacement operator for the number basis as 
\begin{equation}
\hat{P}|n\rangle=|n+1\rangle.
\end{equation}
We write this operator as
\begin{equation}
\hat{P}=\int_0^{2\pi} d\phi e^{-i\phi}\hat{E}(\phi)
\end{equation}
where the states $\hat{E}(\phi)d\phi$ is a projection operator valued measure (POVM) defined by
\begin{equation}
\hat{E}(\phi)=|\phi\rangle\langle \phi|
\label{POVM}
\end{equation}
with
\begin{equation}
|\phi\rangle=\sum_{n=0}^\infty e^{-in\phi}|n\rangle
\label{SG_state}
\end{equation}
This of course is not a physical state. However we will only ever need the POVM $\hat{E}(\phi)$ which is the optimal generalised measurement (POVM) for estimating the phase of a simple harmonic oscillator\cite{SG,Holevo}. 
 
 Define the operator
 \begin{equation}
 \hat{\Phi}=\int_0^{2\pi} d\phi\hat{E}(\phi).
 \end{equation}
 It is then easy to see that
 \begin{equation}
 [ \hat{\Phi},a^\dagger a]=i,
 \end{equation}
so we may regard $\hat{\Phi}$ and $a^\dagger a$ as canonically conjugate pairs. It is now easy to see that the joint operator corresponding to phase difference, $\hat{\Phi}_1-\hat{\Phi}_2$ commutes with the number sum operator $a^\dagger_1 a_1+a^\dagger_2a_2$: a fact which enable one to implement a teleportation protocol based on eigenstates of total number\cite{Coch}. Thus, the operator
\begin{equation} 
\hat{\Phi}_-=(T_1\hat{\Phi}_1-T_2\hat{\Phi}_2)
\end{equation}
with $T_i\equiv 2\pi/\omega_i$, commutes with the total Hamiltonian in Eq.(\ref{tot_ham}) and is accordingly an admissible physical quantity given our (self-imposed) constraint.  Using the assumed values of $\omega_i$ we can write 
\begin{equation} 
\hat{\Phi}_-=T_1\left(\hat{\Phi}_1-\frac 1N\hat{\Phi}_2\right)
\label{phas_dif}
\end{equation}

Now consider the joint phase distribution for the state $|M:N\rangle$, with all the $c_m$ chosen as equal, $c_m=g^{-1/2}$
\begin{equation}
P(\phi_1,\phi_2)={\rm tr}\left (\hat{E}_1(\phi_1)\otimes\hat{E}_2(\phi_2)|M:N\rangle\langle M:N|\right )
\label{phase_dist}
\end{equation}
Using equations Eq.(\ref{POVM}) and Eq.(\ref{SG_state}) we find that
\begin{equation}
P(\phi_1,\phi_2)=\frac{1}{g}\left |\frac{1-e^{ig(N\phi_1-\phi_2)}}{1-e^{i(N\phi_1-\phi_2)} }\right |^2
\end{equation}
When $M \gg N$, or equivalently $g \gg 1$, this is sharply peaked at 
\begin{equation}
N\phi_1-\phi_2=2\pi k,\ \ \ k\ \  \mbox{integer}.
\label{phase_cond}
\end{equation}
Given the definition of $\hat{\Phi}_-$ in Eq.(\ref{phas_dif}), we see that the state $|M:N\rangle$ is a near eigenstate of the phase difference operator. 

The interpretation of Eq.(\ref{phase_cond}) goes as follows. Independent measurements of the phase of each oscillator will reveal a strong correlation. Given an arbitrary ``clock reading" $\phi_1$, the conditional probability of the system's phase $\phi_2$ is sharply peaked around the value $N\phi_1 \bmod 2\pi$. This is a very familiar structure that can be more easily seen using a graphical construction shown in figure \ref{fig1}~a). We end up with the Poincar\'{e} section in the phase space of the  two oscillators for which $\phi_2$ is a function of $\phi_1$.  Classically this immediately implies a dynamical evolution with respect to a coordinate time $t = \phi_1/\omega_1$ that is precisely what one would expect. 

\begin{figure}[h]
\includegraphics[width=6in]{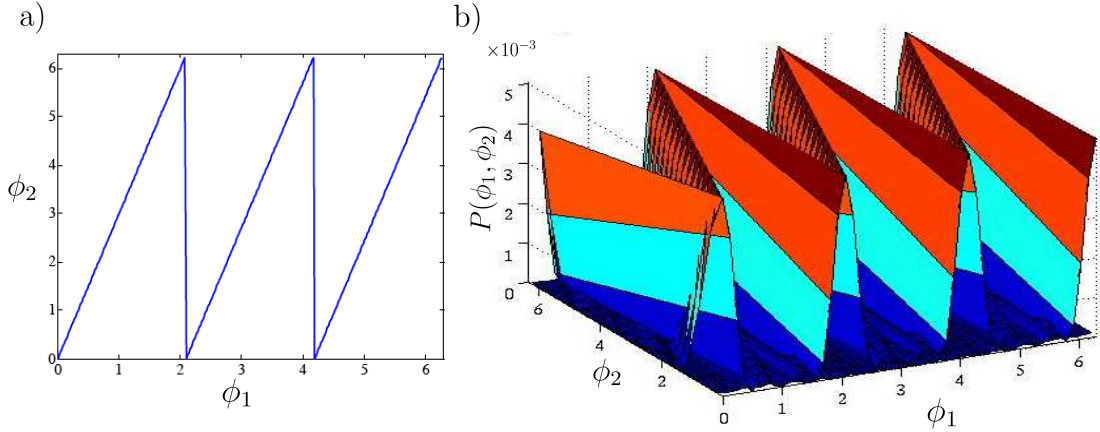}
\caption{a) A schematic representation of the phase space solution for $\phi_1$ and $\phi_2$ in the asymptotic limit $g \rightarrow \infty$, with $N=3$. b) The joint distribution of $\phi_1$ and $\phi_2$ given by Eq.~(\ref{phase_dist}) for the parameters $N = 3$ and $g=14$. As expected, the distribution is sharply peaked around $\phi_2 = N\phi_1 \bmod 2\pi$, so the system undergoes three complete oscillation for $\phi_1 \in [0,2\pi]$.  These correlations inherent in the phases of the two oscillators in a near eigenstate of phase difference.  As $M$ increases, the distribution reaches a delta function and the relation $\phi_2 = \omega_2 t \bmod 2\pi$ becomes exact [as illustrated in figure a)], hence the definition $t = \phi_1/\omega_1$.} \label{fig1}
\end{figure}

However as $M$ is in fact finite, the correlations between $\phi_1$ and $\phi_2$ are imperfect, as seen on figure \ref{fig1}~b). When viewed as a dynamical system and insisting on recovering differential equations that are {\em local} with respect to the ``time" variable $\phi_1$, these fluctuations will be interpreted as noise or error in the frequency of the system.  In other words, quantum fluctuations inherent in the nature of the total state of both oscillators $|N:M\rangle$ will appear as an intrinsic source of noise in the classical record of measured phases.  This  implies an effective decoherence in the local energy eigenstates for the {\em system}, as we now explain.  

Returning to the equivalent expression for $\rho_U$ given in Eq.(\ref{average}) , but now written as a discrete average
\begin{equation}
\rho_U=\lim_{\Omega \rightarrow \infty}\frac{1}{\Omega+1}\sum_{\alpha=0}^\Omega e^{-i\alpha \hat{H}}|\psi\rangle_1\langle\psi|\otimes|\xi\rangle_2\langle\xi|e^{i\alpha \hat{H}}
\label{disc-av}
\end{equation}
we can give a presentation of relational time in terms of a coordinate time $t$. In this picture it is natural to interpret  each term in Eq.(\ref{disc-av}) as the coordinate time evolution with $|\psi\rangle_1\otimes|\xi\rangle_2$ as an {\em initial} state of the two oscillators when they do not interact. We stress that these initial states are completely arbitrary. However from a perspective internal to the universe in which we wish to interpret oscilllator-$1$ as a clock, we may only wish to consider those situations in which $|\psi\rangle_1$ is chosen especially. How is this to be done?

A good clock must have some dynamics, in fact we would like it to evolve rapidly compared to {\em everything else}. This means it must be in a state with a large variance in energy, which here implies a large variance in number $a_1^\dagger a_1$.  So we will take $|\psi\rangle$ so that  
\begin{equation}
\langle \psi|(a_1^\dagger a_1)^2|\psi\rangle-\left (\langle\psi| a^\dagger_1 a_1|\psi\rangle\right )^2\gg 1
\label{good_clock}
\end{equation}
which means that $c_n\equiv\mbox{}_1\langle n|\psi\rangle_1$ is almost independent of $n$ over some large range $K\leq n\leq K+L$ with $L \gg 1$.  In this case, {\em everything else} means simply oscillator-$2$. We do not specify the state of this system at all. Our objective is to recover standard Schr\"{o}dinger  dynamics for this system. 

Suppose we now ask for the conditional state of oscillator-$2$ selected according to all those measurements of $\hat{\Phi}$ on the clock system that gave a particular result $\phi_1$.  This is given by 
\begin{equation}
\rho_2^{(\phi)}={\rm tr}_1\left (|\phi\rangle_1\langle\phi|\otimes I_2\ \rho_U\right )
\end{equation}
Substituting the expression in Eq.(\ref{disc-av}) and using $\hat{H}=a_1^\dagger a_1+Na_2^\dagger a_2$ we find that
\begin{equation}
\rho_2^{(\phi)}=\sum_{\alpha=0}^\Omega{\cal P}(\phi|\alpha)e^{-i\alpha Na_2^\dagger a_2}|\xi\rangle_2\langle \xi|e^{i\alpha N a_2^\dagger a_2}
\label{cond_state}
\end{equation}
where
\begin{equation}
{\cal P}(\phi|\alpha)=\left [p(\phi)\right ]^{-1}\left |\sum_{n=0}^\infty c_n e^{-i(\alpha-\phi)n}\right |^2
\end{equation}
where the normalisation is the inverse of the probability that the result $\phi$ is indeed selected
\begin{equation}
p(\phi)=\frac{1}{\Omega+1} \sum_{\alpha=0}^\Omega {\cal P}(\phi|\alpha).
\end{equation}
The form given in Eq.(\ref{cond_state}) is the same construction as recently proposed by Gambini et al. in their construction of relational time\cite{GPP2004}.  It implies the intrinsic decoherence that we alluded to the the relational construction of the Poincar\'{e} section. 

When $|\psi\rangle_1$ is chosen to be a good clock state according to the prescription Eq.(\ref{good_clock}) we can show that ${\cal P}(\phi_1|\alpha)$ considered as a function of the integer $\alpha$ is sharply peaked at $\alpha=\phi_1$. In this case we may approximate the selected state $\rho_2^{(\phi_1)}$ as 
\begin{equation}
 \rho_2^{(\phi_1)}\approx e^{-iN\phi_1 a_2^\dagger a_2}|\xi\rangle_2\langle\xi| e^{iN\phi_1 a_2^\dagger a_2}
 \end{equation}
 using a simple redefinition of the variable $\phi_1=\omega_1 t$ in terms of a 'time' coordinate and using the fact that $\omega_2/\omega_1=N $ we see that this equation may be written
 \begin{equation}
  \rho_2^{(\phi_1)}\equiv\rho_2(t)\approx e^{-i\omega_2 t a_2^\dagger a_2}|\xi\rangle_2\langle\xi| e^{i\omega_2 t a_2^\dagger a_2}
\end{equation}
  which of course is Schr\"{o}dinger dynamics with respect to coordinate time $t$.  Successive corrections to this approximation describe an intrinsic decoherence in the energy basis. 
  
  To make further progress we need to specify a clock state and compute ${\cal P}(\phi|\alpha)$.  However  a particularly simple model that replaces  unitary Schr\"{o}dinger evolution with a unital semigroup evolution  was first proposed in \cite{GM91}. This is done by taking a Poisson distribution,
 \begin{equation}
{\cal P}(\phi|\alpha)= \frac{(\tilde{\gamma} \phi)^\alpha}{\alpha!}e^{-\tilde{\gamma}\phi}
\end{equation}
where $\tilde{\gamma}$ is dimensionless number which parameterises different clock states. Making the change of variables, $\omega_1 t=\phi,\ \ \   \tilde{\gamma}=\gamma/\omega_1$ we can write this as
\begin{equation}
{\cal P}(t|\alpha) =\frac{(\gamma t)^\alpha}{\alpha!}e^{-\gamma t}
\label{unital_prob}
\end{equation}
where now $\gamma$ may be taken as a new fundamental constant with dimensions of frequency that takes a maximum value and limits us {\em in principle} to a best, but imperfect, clock\cite{Percival}. In the presentation given above, $\gamma$ determines how fast the dynamics of a clock can be by restricting the allowed variance in the energy of the clock to some finite maximum value. One expects that this limit will arise naturally in a fully quantum theory of the universe, that is, one in which both matter and spacetime are given a quantum description. However at this stage it is left as a free parameter. Of course we might expect $(\gamma)^{-1}$ to be related somehow to the Planck time. 

Using Eq.({\ref{unital_prob}) we can easily see that the Schr\"{o}dinger equation is replaced by 
\begin{equation}
\frac{d\rho_2(t)}{dt}=\gamma\left (e^{-ia_2^\dagger a_2/\gamma}\rho_2(t)e^{ia_2^\dagger a_2/\gamma}-\rho_2(t)\right )
\end{equation}
where we see that the right hand side is the generator of a unital semigroup that leaves all quantities that commute with the Hamiltonian invariant in coordinate time.  If, as we expect, $\gamma$ is very large, but not infinite , we can expand in inverse powers of $\gamma$ to obtain a first order correction to Schor\"{o}dinger dynamics
\begin{equation}
\frac{d\rho_2(t)}{dt}=-i[a_2^\dagger a_2,\rho_2(t)]-\frac{1}{2\gamma}[a_2^\dagger a_2,[a_2^\dagger a_2,\rho_2(t)]]+\ldots
\end{equation}
The second term here leads to diffusion\cite{Phase_diff}, in the phase variable $\hat{\Phi}_2$, as suggested in the relational Poincar\'{e} section,  while coherence in the number basis is damped.  The general implications of this dynamics are explored in \cite{GM91}. One important consequence is a modification of the dispersion relations for field theories.

In our construction the emergence of coordinate time, and an intrinsic decoherence process, arises due to the freedom to choose states of the total system as  mixtures of energy eigenstates. This freedom incorporates the same intuition that suggests that the state of the universe must be invariant under the quantum time translation operator. While the universe must be in an energy eigenstate, it is seems likely that the particular eigenvalue realised cannot be known.  A Bayesian perspective would then suggest that we assign the state of the universe as a mixture over energy eigenstates, with an arbitrary distribution. It is this property that enables us to introduce coordinate time as a group average. It would be interesting to consider groups other than the time translation group, for example the Lorentz group. The fact that the mixture is arbitrary gives us the freedom to describe subsystem dynamics in coordinate time from arbitrary initial conditions. The choice of a mixture of energy eigenstates is a departure from what is usually done in canonical  quantum gravity. It would be interesting to revisit the assumptions in canonical quantum gravity to understand this freedom.

\end{document}